\documentclass[12pt]{article}%
\usepackage{amsmath}
\usepackage{natbib}
\usepackage{graphicx}%
\usepackage{amsfonts}%
\usepackage{amssymb}
\usepackage{color}
\usepackage{bm}
\usepackage{setspace}
\usepackage{lineno}

\newcommand{\bbeta}{ \mbox{\boldmath $ \beta $} }

\newcommand{\bZ}{\textbf{Z}}

\newcommand{\bs}{\textbf{s}}

\newcommand{\bX}{\textbf{X}}

\newcommand{\bY}{\textbf{Y}}

\title{Clarifying species dependence under joint species distribution modeling}
\author{Alan E. Gelfand and Shinichiro Shirota\\
Department of Statistical Science, Duke University, NC, USA and\\ AIP Center, RIKEN, Tokyo, Japan}

\begin{document}
\maketitle

\vspace*{0.5cm}

\begin{abstract}

\begin{itemize}
\item[]
\item[1.] Joint species distribution modeling is attracting increasing attention these days, acknowledging the fact that individual level modeling fails to take into account expected dependence/interaction between species.  These models attempt to capture species dependence through an associated correlation matrix arising from a set of latent multivariate normal variables.  However, these associations offer little insight into dependence behavior between species at sites.

\item[2.] We focus on presence/absence data using joint species modeling which incorporates spatial dependence between sites.  For pairs of species, we emphasize the induced odds ratios (along with the joint probabilities of occurrence); they provide much clearer understanding of joint presence/absence behavior.  In fact, we propose a spatial odds ratio surface over the region of interest to capture how dependence varies over the region.   
\item[3.] We illustrate with a dataset from the Cape Floristic Region of South Africa consisting of more than 600 species at more than 600 sites.  We present the spatial distribution of odds ratios for pairs of species that are positively correlated and pairs that are negatively correlated under the joint species distribution model.  
\item[4.] The multivariate normal covariance matrix associated with a collection of species is only a device for creating dependence among species but it lacks interpretation.  By considering odds ratios, the quantitative ecologist will be able to better appreciate the practical dependence between species that is implicit in these joint species distribution modeling specifications.
\end{itemize}
\end{abstract}

\bigskip



\section{Introduction}

Recently, numerous publications on joint species distribution modeling (JSDM) have appeared in the literature \citep{Pollocketal(14), Thorsonetal(15), Ovaskainenetal(16), Clarketal(17)}.  A comparison of such modeling has been presented in Wilkinson2018. Such effort reflects the realization that observation of a community at a site anticipates dependence between the species present at that site. That is, stacked species distribution modeling \citep{GuisanRahbek(11), Calabreseetal(14)}, i.e., modeling the species independently but looking at the results jointly, need not perform well. For example, with presence/absence data, it has been suggested that such modeling may tend to overestimate probability of presence for each species at a site.  Hence the number of presences or the richness at a site \citep{GuisanRahbek(11), Clarketal(17)} may be overestimated.

\cite{GuisanRahbek(11)}, solely in the context of presence/absence data at conceptually point-referenced sites, offer the following criticisms of stacked species distribution modeling: (i) without adding a dispersal filter it may incorrectly predict species in areas that appear environmentally suitable but that are outside their colonizable or historical range \citep{Wiszetal(07)}; (ii) it does not consider any constraints based on the carrying capacity of the local environment which determine the maximum number of species that may co-occur \citep[e.g. species-energy or metabolic theory;][]{Brownetal(04), Currieetal(04)}; and (iii) it does not explicitly consider any rules based on biotic interactions that control species co-occurrences and can exclude species from a community \citep[e.g., through competitive exclusion;][]{Andersonetal(02)}. Due to these three issues, too many species can be predicted to occur in a geographical unit by stacked species distribution models \citep[e.g.,][]{GrahamHijmans(06), PinedaLobo(09)}. Our discussion below will attempt clarification within a probabilistic modeling framework.

Joint species distribution modeling has been developed for presence/absence data, for count (abundance) data, and for composition data \citep{Clarketal(17)}.  Here, we focus primarily on presence/absence data, but offer some extension to abundance data.  Our setting considers say, a large number, $S$, of species and a large number, $n$, of sites.  Site $i, i=1,2,...,n$ provides an $S \times 1$, vector, $\mathbf{Y}_i$ with entries $1$ (presence) or $0$ (absence).   The sites themselves may be viewed, hence modeled, as independent or as spatially dependent, as appropriate.  Regardless, dependence among species, i.e., among the $Y_{ij}$ in $\mathbf{Y}_{i}$, is modeled within a site.

The joint species distribution modeling challenge is the need to model, for each site $i$, the set of $2^{S}$ probabilities associated with the set of possible realizations of $\mathbf{Y}_i$.  Direct modeling of these probabilities requires specification of the $2^S$ probabilities at each site and is clearly infeasible even for relatively small $S$ (while we imagine $S$ of order $10^2$ or more).  The common solution that has been adopted in the literature is to introduce latent variables, $\mathbf{Z}_i$ which drive the responses $\mathbf{Y}_i$.  The $\mathbf{Z}_i$ are modeled as $S \times 1$ multivariate normal vectors which enables tractable model specification though still computationally demanding model fitting.

The contribution here is to note the strong limitation with regard to attempting to understand dependence between pairs of species through the correlations between the corresponding pairs of latent normal variables.  It has been noted, e.g., \cite{Clarketal(17)}, that the $S \times S$ correlation matrix associated  with the latent multivariate normal vectors is only a device for creating dependence among species and that the pairwise correlations lack interpretation.  We demonstrate this convincingly through the use of odds ratios which do offer clear interpretation.  Furthermore, we work with a spatial JSDM, following \cite{Shirotaetal(19)}, which enables, for each pair of species, an odds ratio surface over the study region.  This surface reveals how the dependence between pairs of species varies over the region.

Often, ecologists are more interested in abundance than in presence/absence.  For counts, this would lead us to multivariate (or at least bivariate) spatial Poisson processes.  Such models are ugly to specify and are computationally intractable. However, odds ratios can be extended to \emph{ordinal} categorical data models \citep{Agresti(12)}.  Therefore, if we consider ordinal abundance classifications, e.g., aggregating counts to bins or looking at percent coverage classes, we can extend our ideas to interpretation of dependence in terms of pairwise abundance for species.  We briefly elaborate these ideas below.

In the sequel, we argue the following technical points:\\

(i) We clarify the limitations of independence modeling in terms of capturing behaviors such as allopatry and sympatry.  Further, with presence/absence data, we suggest when stacked SDM's are expected to over-estimate expected richness at sites.  When this is not the case, we can anticipate similar expected richness inference as that for JSDMs but that differences will lie in the associated uncertainties.

(ii) We clarify that the dependence structure associated with the latent multivariate normal model used to drive the JSDM offers little useful information regarding whether joint occurrence of species at a site is encouraged or discouraged.  We demonstrate that odds ratios tell a more useful story in this regard and propose their use for inference.

(iii) We demonstrate that spatial modeling of locations in the context of JSDMs \citep[as in][]{Shirotaetal(19)} adds useful inference regarding dependence.  
For odds ratios, we develop a spatially varying odds ratio surface over a region of interest in order to better understand if and how the nature of pairwise species dependence varies across the region.

We consider a large plant communities dataset from the Cape Floristic Region (CFR) in South Africa \citep[following][]{Shirotaetal(19)}. This dataset
consists of presence-absence measurements for 639 tree species at 662 locations and the spatial JSDM is fitted to this data.  

The format of the paper is as follows.  The second section reminds us of $2 \times 2$ tables with associated dependence.  Section 3 briefly considers species richness.  Section 4 reviews the basics of JSDMs, including a minor technical issue.  It also supplies connection to odds ratios.  Section 5 brings in formal spatial modeling while Section 6 presents the results for the foregoing CFR dataset.  Section 7 considers species dependence with abundance data and Section 8 offers a brief conclusion.

\section{$2 \times 2$ tables and familiar ecological dependence notions}

We review the role of $2 \times 2$ tables for learning about pairwise species dependence.  In particular, these tables facilitate quantification of basic notions such as sympatry and allopatry.
For illustration, we begin with a single site and two species in which case a joint model requires specification of four probabilities, $(p_{00}, p_{01}, p_{10}$, and $ p_{11})$  which sum to 1.  The subscripts indicate absence (0) or presence (1) of the first and second species, respectively.  Customarily the probabilities are presented in a $2 \times 2$ table (where the rows are associated with species 1 and the columns are associated with species 2),

\bigskip

\begin{tabular}{|c|cc|c|}
   & 0 & 1 &  \\ \hline
  0 & $p_{00}$ & $p_{01}$ & $p_{0.}$ \\
  1 & $p_{10}$ & $p_{11}$ & $p_{1.}$ \\ \hline
   & $p_{.0}$ & $p_{.1}$ & 1 \\
\end{tabular}

\bigskip

For instance, the probabilities in the following table would provide an example of strong sympatry, i.e., when the species are present, there is a strong chance that they are present together.

\bigskip

\begin{tabular}{|c|cc|c|}
   & 0 & 1 &  \\ \hline
  0 & $.14$ & $.02$ & $.16$ \\
  1 & $.04$ & $.80$ & $.84$ \\ \hline
   & $.18$ & $.82$ & 1 \\
\end{tabular}

\bigskip

Formally, sympatry is characterized by encouraging joint occurrence or joint absence, e.g.,
$$P(\text{species 2  present}|\text{species 1 present}) > P(\text{species 2 present}).$$
With the notation above, we have $\frac{p_{11}}{p_{1.}} > p_{.1}$.  Switching the species, we have $\frac{p_{11}}{p_{.1}} > p_{1.}$.  So, in either case, $p_{11} > p_{1.}p_{.1}$, a departure from independence.  Expressed through odds, we have $\frac{p_{11}}{p_{10}} > \frac{p_{01}}{p_{00}}$, the odds for presence of species 2 when species 1 is present are greater than the odds when species 1 is not present.   Vice versa, $\frac{p_{11}}{p_{01}} > \frac{p_{10}}{p_{00}}$, the odds for presence of species 1 when species 2 is present are greater than the odds when species 2 is not present. Either expression implies that the odds ratio, $\theta \equiv \frac{p_{11}p_{00}}{p_{10}p_{01}} >1$, is capturing positive dependence, since under independence, the odds ratio always equals $1$.  For the table above, $\theta = 140 $, very large, far above $1$; we have very strong positive dependence between the species.   Odds ratios are often presented on the log (base $10$) scale, placing them on the whole real line with $0$ as the center.  That is, the log odds ratio is employed in order to ``symmetrize'' departure from independence around $0$.  Here, log$\theta = 2.15$.

Next, consider the table,

\bigskip

\begin{tabular}{|c|cc|c|}
   & 0 & 1 &  \\ \hline
  0 & $.03$ & $.49$ & $.52$ \\
  1 & $.47$ & $.01$ & $.48$ \\ \hline
   & $.5$ & $.5$ & 1 \\
\end{tabular}

\bigskip

Such a table captures very strong allopatry.  If one species is present, with very high probability the other is not.  Formally, allopatry is characterized by discouraging co-occurrence, e.g.,
$$P(\text{species 2  present}|\text{species 1 present}) < P(\text{species 2 present}).$$
With the notation above, we have $\frac{p_{11}}{p_{1.}} < p_{.1}$.  Switching the species, we have $\frac{p_{11}}{p_{1.}} < p_{.1}$ so in either case, $p_{11} < p_{1.}p_{.1}$, again a departure from independence.  Expressed through odds, we have $\frac{p_{11}}{p_{10}} < \frac{p_{01}}{p_{00}}$.  The odds for presence of species 2 when species 1 is present are less than the odds when species 1 is not present.   Vice versa, $\frac{p_{11}}{p_{01}} < \frac{p_{10}}{p_{00}}$, the odds for presence of species 1 when species 2 is present are less than the odds when species 2 is not present. Now, either expression implies that the odds ratio, $\theta = \frac{p_{11}p_{00}}{p_{10}p_{01}} <1$, capturing negative dependence.  For the table above, $\theta = .0013$, very small, well below $1$; we have very strong negative dependence between the species. Also, log$\theta = -2.89$.

In summary, what the independence model gets wrong but the joint species distribution helps with is specifying the joint probabilities.  That is, at a given site, the joint probability that species $j$ and $j'$ co-occur, $p_{11}^{(j,j')}$ will be poorly estimated by an estimate of $p^{(j)}p^{(j')}$.  Hence, this will be true for the other joint probabilities.

Accordingly, we suggest that the odds ratio provides a useful tool for learning about species dependence with regard to presence/absence.  Below, we propose to employ odds ratios to assess to departure from independence, with clear interpretation in terms of encouraging or discouraging joint occurrence or joint absence for pairs of species.

A further remark is perhaps useful here.  Under a JSDM we typically consider that many possible species can potentially occur at a site, e.g., $O(10^2)$, as in our dataset.  However, in practice, typically we find just a few species at a given site, say, perhaps at most $10$.  As a result, at site $\bs$, with $1_{j}(\bs)$ denoting the indicator function indicating presence or absence of species $j$ at location $(\bs)$, most of the $P(1_{j}(\bs)=1) \equiv p_{j}(\bs) \approx 0$.  Therefore, most of the $p_{ab}^{(jj')}(\bs), a,b = 0,1$ are $ \approx 0$.  Fortunately, the odds ratio is insensitive to rescaling the four probabilities.  If we re-normalized the four probabilities for species $j$ and $j'$ in their associated $2 \times 2$ table to sum to $1$, the odds ratio is unaffected.

In the sequel, we will focus on species in pairs, using the odds ratio.  This is analogous to specifying correlation pairwise in providing multivariate normal distributions.  Perhaps most importantly, since independence modeling underlies stacked species distribution models, such models will not be able to capture sympatric or allopatric behavior for pairs of species.

\section{Species richness}

In characterizing species distributions at sites, a common measure is species richness.  Species richness records the number of distinct species present at the site.  Again, if $1_{j}(\bs)$ is the indicator of presence of species $j$ at site $\bs$, then the observed richness is $\texttt{Rich}(\bs) = \sum_{j=1}^{S} 1_{j}(\bs)$.  (Just to be clear, $\{1_{j}(\bs)\}$ does not constitute a multinomial trial but, rather, a set of dependent Bernoulli trials.) Note that, regardless of whether we model species independently, e.g., using stacked species distribution models or dependently, using JSDM's, $E(\texttt{Rich}(\bs)) = \sum_{j=1}^{S}E(1_{j}(\bs)) = \sum_{j=1}^{S}P(1_{j}(\bs) = 1)$.  So, if we model species independently, we need not expect to incorrectly estimate expected richness.

This remark merits a bit of amplification.  The issue here is whether the marginal expectations across the species under a joint model will tend to agree with the corresponding expectations across species under independent models.  Because the joint model considers the data for all of the species at a site while the individual models consider the data only for the individual species at the site, unconstrained by the overall presence/absence at the site, intuitively, we might anticipate the latter expectations to be larger.  This would suggest predicting higher richnesses using a stacked species distribution model, in support of the discussion in the Introduction.  However, there is no formal theory to support this intuition and, omitting details, maximum likelihood estimation will produce expectations that agree.

However, we expect to incorrectly estimate uncertainty in richness when the sum of the indicator variables is not a sum of independent variables.  That is, richness does not reflect the chance of joint presence or absence.  Formally, Var$(\texttt{Rich}(\bs)) = \text{Var}(\sum_{j=1}^{S} 1_{j}(\bs)) = \sum_{j=1}^{S} \text{Var}(1_{j}(\bs)) + \sum_{j < j'} \text{Cov}(1_{j}(\bs), 1_{j'}(\bs))$.  However, $\text{Cov}(1_{j}(\bs), 1_{j'}(\bs)) = p_{11}^{(j,j')}(\bs) - p_{j}(\bs)p_{j'}(\bs)$.  We can see that departure from independence can affect this term and there how it can affect the variance in observed richness.  Again, since the odds ratio provides pairwise information about departure from independence, we can identify where uncertainty in richness is affected.  Specifically, the presence of sympatric or allopatric behavior for pairs of species influences the uncertainty in species richness.

\section{Joint species distribution modeling}

We offer a brief review of joint species distribution modeling in the nonspatial case.   We take our data to be $Y_{ij}$ a binary variable indicating presence ($Y_{ij}=1$) or absence ($Y_{ij}=0$) of species $j$ at site $i$. Following either \cite{Clarketal(17)} or \cite{Ovaskainenetal(16)}, we associate with each $Y_{ij}$ a latent normal variable $Z_{ij}$.  In fact, we have a latent multivariate normal vector $\bZ_{i}$ which yields the observed $\bY_{i}$.  The question to ask is how shall we connect the $Z$'s to the $Y$'s?

We adopt notation in the spirit of \cite{Ovaskainenetal(16)}, letting $L^{F}_{ij}$ and $L^{R}_{ij}$ denote the fixed and random effects contributions, respectively, which are included additively in the modeling of $Z_{ij}$.  More will be said about the forms of these $L$'s below but $L$ itself is intended to suggest a linear form.  As the definitions suggest, we will view $L^{F}_{ij}$ as a nonrandom component in the specification for $Z_{ij}$. It will have parameters - regression coefficients - in it and, so, in a Bayesian hierarchical modeling framework, these coefficients would be viewed and modeled as random.  We will view $\mathbf{L}^{R}_{i}$ as a multivariate normal random variable with mean $\mathbf{0}$ and dependence structure given by an $S \times S$ correlation matrix, $H$.  So, marginally, $L^{R}_{ij} \sim N(0,1)$.

Under a functional relationship between $Y_{ij}$ and $Z_{ij}$, $Y_{ij}=1(Z_{ij} >0)$ and $P(Y_{ij}=1) = P(Z_{ij}>0)$.  Consider the following two specifications for $Z_{ij}$:
\begin{itemize}
    \item[(i)] $Z_{ij} = L^{F}_{ij} + L^{R}_{ij}$
    \item[(ii)] $Z_{ij} = L^{F}_{ij} + L^{R}_{ij} + \epsilon_{ij}$
\end{itemize}
where, in (ii), the $\epsilon_{ij}$ are pure error terms, i.e.,  independent and identically distributed  normal random variables with mean $0$ and variance $1$. Then, under (i),  given $L^{F}_{ij}$, the $Z_{ij}$ are dependent and, with $\Phi$ denoting the standard normal cumulative distribution function,
\begin{equation}
P(Y_{ij}=1) = P(Z_{ij}>0) = \Phi(L^{F}_{ij}).
\label{eq:fixed}
\end{equation}
Under (ii), given $L^{F}_{ij}$ and $L^{R}_{ij}$, the $Z_{ij}$ are independent and
\begin{equation}
P(Y_{ij}=1) = P(Z_{ij}>0) = \Phi(L^{F}_{ij}+ L^{R}_{ij}).
\label{eq:random}
\end{equation}
Now, the $\Phi$'s are dependent.

Viewing the sites as independent, we are modeling solely dependence of species within sites. If we adopt (ii) above, we model $P(Y_{ij}=1)$ to include both fixed and random effects.  Dependence is introduced through the specification for $L^{R}_{ij}$.  In fact, we create $\text{corr}(L^{R}_{ij}, L^{R}_{ij'}) =H_{jj'}$.

Two remarks are worth making.  First, under (ii), dependence between species is only introduced in the probability of presence, the so-called second stage of a hierarchical model, as we see from (2).  In fact, it is the correlation between $\Phi^{-1}(P(Y_{ij}=1))$ and $\Phi^{-1}(P(Y_{ij'}=1))$.

Second, we see clearly that stochastic dependence between probability of presence replaces modeling of interaction \citep{Clarketal(17)}.  As noted in the Introduction, all such JSDM's concede that modeling of interactions between species at a site is too demanding, that probabilistic dependence is the most tractable surrogate.  Again, we have the question of what such correlation means.  In this regard, it is associated with \emph{residuals} as (ii) reveals, i.e., adjusted for the \emph{mean}, $L^{F}_{ij}$.  

Moreover, at any site, we will find only a small subset of the $S$ species present.  That is, $\mathbf{Y}_i$ will be predominantly comprised of $0$'s.  Nonetheless, we create pairwise associations for all pairs of species and these associations do not vary with site.  So, it is evident that these associations have little to do with the actual realization of $\mathbf{Y}_{i}$ at site $i$.  

Furthermore, is a positive association suggestive of co-occurrence or of a potential substitution effect, i.e.,  a particular species is present but another, say similar one, could equally well have been successful there?  This leads to discussion presented in \cite{ZobelAntos(97)} and in \cite{Ovaskainenetal(16)} regarding the global species pool (all existing species), the regional species pool (those able to colonize an area), and the local species pool (those found at the finest scale considered).  Recalling our discussion above, these concerns lead us to consider odds ratios in order to enhance our understanding of pairwise dependence for species.

Alternatively, suppose we take model (i) for $Z_{ij}$?  This moves the modeling of dependence to the first stage specification.  Now the $Z_{ij}$ are dependent and therefore, so are the $Y_{ij}$.  This direct dependence approach is advocated in \cite{Clarketal(17)}.  Now, $\text{corr}(Z_{ij}, Z_{ij'}) = \text{corr}(L^{R}_{ij}, L^{R}_{ij'}) = H_{jj'}$.  The concern here is that now the probability of presence has no random effects in it; simply, $P(Y_{ij}=1) = \Phi(L^{F}_{ij})$ is entirely driven by covariates.  We have a basic probit regression for every species. Moreover, we have a recurring issue: how do we usefully interpret a correlation between normal random variables with regard to the association between the binary variables, $Y_{ij}$ and $Y_{ij'}$?


Now, consider a conditional specification, $[Y_{ij}|Z_{ij}]$.  Under a probit link function, $P(Y_{ij}=1) = \Phi(Z_{ij})$.  So, under (i), we obtain $\Phi(L^{F}_{ij}+ L^{R}_{ij})$, the form in \cite{Ovaskainenetal(16)}.  We can conclude that using (ii) under a functional specification or (i) under a conditional specification produces the same probability of presence.  Therefore, if our goal is merely to obtain $\Phi(L^{F}_{ij}+ L^{R}_{ij})$ as the probability of presence, we can achieve this under either specification.  However, if the functional specification is used, we must adopt (ii) above for the $Z$'s.  This distinction seems muddled in, e.g., \cite{Wilkinsonetal(18)}.  In any event, in the sequel, we work exclusively with the functional specification under (ii).

%

\subsection{Connection to odds ratios}


For the JSDMs referenced above, again, dependence across species is captured through the pairwise correlation between species in the latent bivariate normal distribution which yields the pair of binary responses.  However, under the JSDM, we do not model the $p_{a,b}^{(j,j')}, a,b = 0,1$ directly but, rather, we model the parameters in the latent multivariate normal distribution and, as a result, each of these probabilities is a smooth function of these parameters.

A key technical point we make here is the fact that there is no direction connection between say $\rho^{(j,j')}$ and the odds ratio associated with the induced $2 \times 2$ table of joint probabilities for the species pair, $j,j')$ at site $i$.  Specifically, suppose the latent bivariate normal has mean $\left(
                                                \begin{array}{c}
                                                  \mu_{i}^{(j)} \\
                                                  \mu_{i}^{(j')} \\
                                                \end{array}
                                              \right)$ and correlation matrix $\left(
                                                                                \begin{array}{cc}
                                                                                  1 & \rho^{(j,j')}  \\
                                                                                  \rho^{(j,j')} & 1 \\
                                                                                \end{array}
                                                                              \right)$.
Then,
\begin{equation}
\theta_{i}^{(j,j')} = \frac{p_{i,00}^{(j,j')}p_{i,11}^{(j,j')}}{p_{i,10}^{(j,j')}p_{i,01}^{(j,j')}} = \frac{P(Z_{ij}<0, Z_{ij'}<0)P(Z_{ij}\geq 0, Z_{ij'} \geq 0)}{P(Z_{ij} \geq 0, Z_{ij'}<0)P(Z_{ij}<0, Z_{ij'} \geq 0)}.
\label{eq:odds}
\end{equation}
Direct examination of the expressions for the double integrals needed in \eqref{eq:odds} shows that each probability is a function of $\mu_{i}^{(j)}$, $\mu_{i}^{(j')}$, and $\rho^{(j,j')}$.  The appendix provides the attractive result that $\theta_{i}^{(j,j')}$ is non-decreasing in $\rho^{(j,j')}$ for fixed $\mu_{i}^{(j)}$ and $\mu_{i}^{(j')}$.  However, the latent correlations do not readily determine the strength/magnitude of the odds ratios, particularly in the presence of  $\mu_{i}^{(j)}$, $\mu_{i}^{(j')}$.  Specifically, the odds ratios have direct ecological interpretation while the correlations do not.  Altogether, it seems that correlation matrix $H$ offers little insight into whether pairs of species are encouraged or discouraged for joint occurrence.

\begin{figure}[ht]
 \begin{minipage}{0.48\hsize}
  \begin{center}
   \includegraphics[width=8cm]{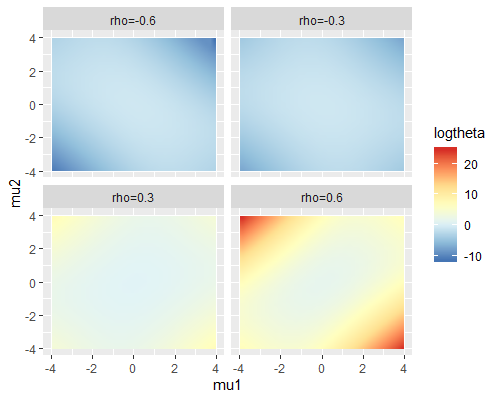}
  \end{center}
 \end{minipage}
 \hfill
 \hfill
 \begin{minipage}{0.48\hsize}
  \begin{center}
   \includegraphics[width=8cm]{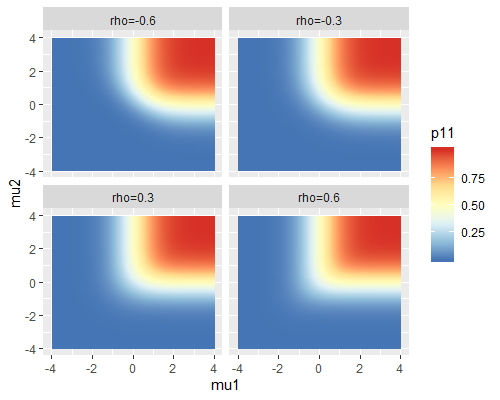}
  \end{center}
\end{minipage}
  \caption{$\log \theta$ (left) and $p_{11}$ (right) with respect to $\rho$ and $\mu$'s}
  \label{fig:ex1}
\end{figure}

Figure \ref{fig:ex1} shows the log odds ratio ($\log \theta$) and the joint occurrence probability ($p_{11}$) as a function of $\rho$ for pairs of $\mu$'s. The monotonically increasing trend is observed in both $\log \theta$ and $p_{11}$ with respect to $\rho$.  Note that, according to the $\mu$'s, we can have $p_{11}$ arbitrarily large with $\rho <0$. That is, $\rho$ supplies essentially no information regarding the probability of joint occurrence of the species. (A similar argument can be made in terms of joint absence.)  The appendix further argues that we can not have contradictory signs for $\rho$ and log$\theta$.  If $\rho >0$, log$\theta \geq 0$, if $\rho <0$, log$\theta \leq 0$.  However, again according to the $\mu$'s, for a given $\rho >0$, we can have log$\theta$ as large or small a positive value as wish.  For a given $\rho <0$, we can have log$\theta$ as large or small a negative value as wish.  So, again, $\rho$ provides little indication of the strength of dependence between species.  


%

To elaborate the inference further, working within a Bayesian framework, analysis of odds ratios and joint occurrence probabilities is a post-model fitting activity.  We can examine any pair of species at any location since these objects are functions of the model parameters; posterior samples of model parameters provide posterior samples of these objects.  As a result, we can obtain the posterior distribution for say, $\theta_{i}^{(j,j')}$.  We can summarize this distribution with a posterior mean but, we can also calculate probabilities such as $P(\theta_{i}^{(j,j')} >1|\text{data})$.  

To be more explicit regarding how this is implemented, under Markov chain Monte Carlo model fitting, we obtain posterior samples, say $\mu_{i,b}^{(j)}, \mu_{i,b}^{(j')}, \rho_{b}^{(j,j')}, b=1,2,...,B$.  Each term in \eqref{eq:odds} is a double integral which is a function of $(\mu_{i}^{(j)}, \mu_{i}^{(j')}, \rho^{(j,j')})$.  So, each sample, $\mu_{i,b}^{(j)}, \mu_{i,b}^{(j')}, \rho_{b}^{(j,j')}$ produces a posterior realization of each of the four terms on the right side of \eqref{eq:odds}, e.g., $p_{i,00,b}^{(j,j')}$, hence a posterior realization, $\theta_{i,b}^{(j,j')}$.  Across $b=1,2,...,B$, we obtain a posterior sample of size $B$ for each of the terms on the right side of \eqref{eq:odds} as well as the induced odds ratio.   The double integrals that are needed can be computed using approximations or numerically.

A fully Monte Carlo alternative is to generate many $(Z_{ij}, Z_{ij'})$ pairs, hence many $(Y_{ij}, Y_{ij'})$ pairs for each $\mu_{i,b}^{(j)}, \mu_{i,b}^{(j')}, \rho_{b}^{(j,j')}$.  The collection can be placed into a $2 \times 2$ table from which we can obtain a posterior realization of each of the probabilities on the right side of \eqref{eq:odds} as well the odds ratio.

In fact, if we work with the functional JSDM and specification (ii) above, then $Z_{ij}$ and $Z_{ij'}$ are conditionally independent given $L^{F}_{ij}$, $L^{F}_{ij'}$, $L^{R}_{ij}$ and $L^{R}_{ij'}$.  So, we only need to calculate univariate cumulative normal distribution functions.

\subsection{A useful digression}

It may be helpful to distinguish $\theta_{i}^{(j,j')}$ in \eqref{eq:odds} from a different odds ratio which is frequently examined in the literature, $\gamma_{i}^{(j,j')} = \frac{p_{ij}}{1-p_{ij}}\div \frac{p_{ij'}}{1-p_{ij'}}$.  The latter is an odds ratio associated with homogeneity, not independence.  We can make a connection to logistic regression for presence/absence with marginal rather than joint probabilities. That is, suppose we seek to compare the log odds ratio at site $i$  for species $j$ vs species $j'$.  If logit$\frac{p_{ij}}{1-p_{ij}} = \bX_{i}\bbeta_{j}$, then $\text{log}\gamma_{i}^{(j,j')} = \bX_{i}(\bbeta_{j} - \bbeta_{j'})$ and homogeneity is equivalent to $\bbeta_{j} = \bbeta_{j'}$.  So, here homogeneity is interpreted as clustering the pair of species.  In the calculation of $\gamma_{i}^{(j,j')}$, joint modeling of species $j$ and $j'$ is not considered.  The analogue with probit modeling would yield $\Phi^{-1}(p_{ij}) - \Phi^{-1}(p_{ij'}) = \bX_{i}(\bbeta_{j} - \bbeta_{j'})$.

Extending to the notation above, using equations \eqref{eq:fixed} and \eqref{eq:random}, we have $\Phi^{-1}(p_{ij}) = L_{ij}^{F}$ and $\Phi^{-1}(p_{ij}) = L_{ij}^{F} +L_{ij}^{R}$, respectively.  Now, $\Phi^{-1}(p_{ij}) - \Phi^{-1}(p_{ij'})= L_{ij}^{F} - L_{ij'}^{F}$ and $L_{ij}^{F} - L_{ij'}^{F} + L_{ij}^{R} - L_{ij'}^{R}$, respectively.  In the latter case, with similar modeling for the logits, $\text{log}\lambda_{i}^{(j,j')}$ becomes a random variable, i.e., a function of the random effects of the form $L_{ij}^{R} - L_{ij'}^{R}$.

\section{Spatial dependence considerations}

Next, suppose we bring in space and spatial dependence.  We can capture spatial explanation through spatially referenced predictors (covariates) to build a spatial regression.  We refer to this as first order spatial modeling.  However, in addition, we wish to bring in spatial dependence in the sense that locations closer to each other are anticipated to provide similar probability of presence for a species (adjusted for covariates).   This is referred to as second order dependence and is captured through spatial random effects.  That is, we add random effects to the structure, customarily assumed to come from a Gaussian process.  These random effects are usually modeled as a smooth (mean square continuous) realization of the process and will tend to be more similar at locations closer to each other in space.  This is accomplished with continuous responses using so-called geostatistical models \citep{BanerjeeCarlinGelfand(14), CressieWikle(11)}.  With binary responses, they are usually introduced through a latent Gaussian process in one of two ways, as we clarify below.

It is critical to distinguish the two types of dependence in play here.  The first is the dependence with regard to presence/absence among species at a site while the second is the dependence in presence/absence across sites.  The latter leads to the notion of cross-covariance, i.e., for the latent Gaussian variables, the covariance between say $Z_{j}(\mathbf{s})$ and $Z_{j'}(\mathbf{s}')$.  In turn, this leads to the dependence between presence of species $j$ at location $\mathbf{s}$ and species $j'$ at location $\mathbf{s}'$.  See \cite{BanerjeeCarlinGelfand(14)} for a full discussion of cross covariance specification and see the paper \cite{Shirotaetal(19)} for joint species distribution modeling, taking into account spatial dependence between locations.

So, in the foregoing notation the independent random effects will be replaced by spatially dependent random effects and we modify notation above by attaching location $\mathbf{s}_i$ to site $i$ and writing $Y_{ij} \equiv Y_{j}(\mathbf{s}_{i})$.  Now we envision point level modeling with a conceptual presence/absence variable, $Y_{j}(\mathbf{s})$ for species $j$ at every location, $\mathbf{s}$, in a study region of interest, say $D$.  As a result, we have a realization of a presence/absence surface for species $j$, $\{Y_{j}(\mathbf{s}): \mathbf{s} \in D\}$.  This surface is observed at $\{\mathbf{s}_{i}, i=1,2,...,n\}$.

It is worth remarking that we are not considering presence/absence associated with areal units, i.e., whether there is at least one individual of species $j$ in the areal unit.  This leads to very different modeling for presence/absence (and is beyond the scope of consideration here). See \cite{GelfandShirota(19)} for careful discussion and development of the differences and see \cite{Shirotaetal(19)} for full development of JSDM's for the point-referenced spatial case.

With regard to the $Z$'s, extending Section 4, now we have:
\begin{itemize}
    \item[(i)] $Z_{j}(\mathbf{s}) = L^{F}_{j}(\mathbf{s}) + L^{R}_{j}(\mathbf{s})$ or
    \item[(ii)] $Z_{j}(\mathbf{s}) = L^{F}_{j}(\mathbf{s}) + L^{R}_{j}(\mathbf{s}) + \epsilon_{j}(\mathbf{s})$
\end{itemize}
Suppose $L^{F}_{j}(\mathbf{s})$ is a continuous surface over $D$ (if defined through regressors, then the regressors need to be continuous over $D$) and suppose $L^{R}_{j}(\mathbf{s})$ is a realization of a Gaussian process \citep{BanerjeeCarlinGelfand(14)} which produces mean square continuous realizations. (A sufficient condition is that the correlation function of the Gaussian process be continuous at $\mathbf{0}$.)  Then, under (i), $Z_{j}(\mathbf{s})$ is a continuous surface while under (ii) $Z_{j}(\mathbf{s})$ is everywhere discontinuous because $\epsilon_{j}(\mathbf{s})$ is.

Again, consider the functional specification, now $Y_{j}(\mathbf{s}) = 1(Z_{j}(\mathbf{s}) >0)$  (which is referred to as a clipped Gaussian field in the literature \citep[e.g.,][]{DeOliveira(00)}, and the conditional specification, now $[Y_{j}(\mathbf{s})|Z_{j}(\mathbf{s})]$.  Suppose, we work with a probability of presence surface which is the analogue of that in Section 4, i.e., $P(Y_{j}(\mathbf{s})=1) = \Phi(L^{F}_{j}(\mathbf{s}) + L^{R}_{j}(\mathbf{s}))$.  Then, following the previous paragraph, for species $j$, the probability of presence surface is continuous over $D$.  However, under the conditional specification, each $Y_{j}(\mathbf{s})$ is drawn as a conditionally independent Bernoulli variable given its probability of presence.  Hence, the realized presence/absence surface, $\{Y_{j}(\mathbf{s}): \mathbf{s} \in D\}$ here is everywhere discontinuous.  This seems unsatisfying; the realized presence/absence surface should manifest \emph{local} smoothness.  It should exhibit local subregions where it is $0$ and local subregions where it is $1$.

For the functional specification, under (ii), since $Z_{j}(\mathbf{s})$ is everywhere discontinuous, we can not obtain local continuity for the $Y_{j}(\mathbf{s})$ surface.  However, under (i), if the $Z_{j}(\mathbf{s})$ is continuous, with the functional specification, we can obtain local continuity for the $Y_{j}(\mathbf{s})$ surface.

The point here is that, with spatial modeling, if we value local smoothness in the realized presence/absence surface, if we think that such smoothness more appropriately captures real world behavior of process realizations, then we should work with the functional specification since this smoothness can never be achieved with the conditional specification.

Turning to the joint modeling, we now envision site specific two way tables at each $\bs$, comprised of $p_{ab}^{(j,j')}(\bs)$  ($a,b,=0,1$), resulting in a ``surface'' of such tables which can be summarized by a log odds ratio, $\text{log}\theta(\bs)^{(j,j')}$, surface.  The spatial dependence modeling provides smoothing for these surfaces as well as interpolation of these surfaces to unobserved locations.

In terms of inference, the JSDM modeling focuses on a large number of species and a large number of (potentially) spatially dependent sites \citep{Clarketal(17), Ovaskainenetal(16)}. In this regard, interest is in clustering and dimension reduction in order to facilitate computation.  Below, we fit such a spatial model to the dataset mentioned in the Introduction.   Then, we summarize for a subset of five species to illustrate the odds ratio story.  The primary benefit in fitting a much bigger model is that we obtain very small but more realistic cell probabilities for most cells in the pairwise $2 \times 2$ tables.

Because the story here is more concerned with pairwise inference for species behavior, we could effectively illustrate the utility of odds ratios using a model with only a small number of illustrative species.  In any event, the key point is that we extend the earlier marginal spatial work which presents individual probability of presence surfaces for each of a set of species \citep{Latimeretal(06), Gelfandetal(05)}. Now, we focus on pairwise probability of joint occurrence, of joint absence of species, and on pairwise log odds ratios, again as surfaces over the region of interest.

\subsection{Model and inference details}


We briefly present the spatial JSDM for presence/absence data from \cite{Shirotaetal(19)}. Let $\mathcal{D}\subset \mathbb{R}^{2}$ be a bounded study region with $\mathcal{S}=\{\bm{s}_{1}, \ldots, \bm{s}_{n}\} \in D$ a set of plot locations, and $\bm{Z}_{i}:=\bm{Z}(\bm{s}_{i})\in \mathbb{R}^{S}$ be an $S\times 1$ latent vector of continuous variables at location $\bm{s}_{i}$.
Under independence for the locations, the model for $\bm{Z}_{i}$ is specified as
\begin{align}
\bm{Z}_{i}=\mathbf{B}\bm{x}_{i}+\bm{\epsilon}_{i}, \quad \bm{\epsilon}_{i}\stackrel{iid}{\sim} \mathcal{N}_{S}(\bm{0}, \mathbf{\Sigma}), \quad \text{for} \quad i=1,\ldots,n \label{eq:(2.1)}
\end{align}
where $\mathbf{B}$ is an $S\times p$ coefficient matrix, $\bm{x}_{i}$ is a $p\times 1$ covariate vector at location $\bm{s}_{i}$ and $\mathbf{\Sigma}$ is a $S\times S$ covariance matrix for species.
This model has $\mathcal{O}(S^2)$ parameters, $S(S+1)/2$ parameters from $\mathbf{\Sigma}$ and $pS$ parameters from $\mathbf{B}$.
For example, for $S=300$ species and $p=3$ covariates, the model contains $46,050$ parameters.

\cite{Taylor-Rodriguezetal(17)} propose a dimension reduction approximation to $\mathbf{\Sigma}$ that allows the number of parameters to grow linearly in $S$.
They approximate $\mathbf{\Sigma}$ with $\mathbf{\Sigma}^{*}=\mathbf{\Lambda}\mathbf{\Lambda}^{T}+\sigma_{\epsilon}^2\mathbf{I}_{S}$ and replace the above model with
\begin{align}
\bm{Z}_{i}=\mathbf{B}\bm{x}_{i}+\mathbf{\Lambda}\bm{w}_{i}+\bm{\epsilon}_{i}, \quad \bm{\epsilon}_{i}\sim \mathcal{N}_{S}(\bm{0}, \sigma_{\epsilon}^2 \mathbf{I}_{S}), \quad \text{for} \quad i=1,\ldots,n \label{eq:(2.2)}
\end{align}
where the random vectors $\bm{w}_{i}$ are i.i.d. with $\bm{w}_{i}\sim \mathcal{N}_{r}(\bm{0}, \mathbf{I}_{r})$ and $\mathbf{\Lambda}$ is an $S\times r$ matrix with $r\ll S$.
Now, $\mathbf{\Sigma}^{*}$ has only $Sr+1$ parameters, the estimation problem of $\mathcal{O}(S^2)$ parameters is reduced to that of $\mathcal{O}(S)$ parameters.  We refer to this specification as the dimension reduced nonspatial model.

According to (\ref{eq:(2.2)}), the $\bm{Z}_{i}$ are conditionally independent given $\bm{B}$ and $\mathbf{\Lambda}$, i.e., the  $\bm{w}_{i}$ are independent across locations.
However, envisioning spatial dependence for  plot locations that are relatively close each other, we introduce spatial dependence into $\bm{w}_{i}$ to enable us to improve the prediction for new plot locations in the study region. That is, suppose we envision the $r \times 1$ vector $\bm{w}_{i}$ as $\bm{W}(\bm{s}_{i})$, a realization of an $r$ dimensional Gaussian process at $\bm{s}_{i}$.  In fact, we let the components of the vector be associated with independent and identically distributed Gaussian processes with common exponential covariance function.
We refer to this modeling specification as the dimension reduced spatial model. Again, \cite{Taylor-Rodriguezetal(17)} consider the entries in $\mathbf{W}^{(h)}$ to be independent across $i$ (i.e., across sites) while we introduce spatial dependence across $i$ through a GP for each column of $\mathbf{W}$.  The full model is more complex than we have presented here, including the use of Dirichlet process specifications as part of the dimension reduction to cluster species.  The interested reader is encouraged to consult \cite{Shirotaetal(19)} for explicit details as well as model fitting and prediction details.  

As a brief aside, with regard to modeling the spatial dependence structure, in principle, each species might have its own spatial range/decay parameter.  However, under the dimension reduction we can include at most $r < < S$ decay parameters.  So, the issue is whether incorporating a common decay parameter for the latent GP's, i.e.,  a separable model, will sacrifice much compared with employing $r$ decay parameters when $r$ is say 3 to 5.  The implications for the species level spatial dependence behavior are expected to be negligible.

For binary response data in the form of presence-absence, a logit or probit model specification is often assumed. The data-augmentation algorithm proposed by \cite{Chib(98)} for multivariate probit regression, which improves the mixing of the Markov chain Monte Carlo (MCMC) algorithm, is adapted.
It is the functional relationship form described above in Section 4,
\begin{align}
Y_{i}^{(j)}=\begin{cases}
             1 & Z_{i}^{(j)}\ge 0 \\
             0 & Z_{i}^{(j)}< 0 \\
            \end{cases}, \quad \text{for} \quad j=1,\ldots, S, \quad i=1,\ldots,n. \label{eq:(3.1)}
\end{align}

\section{The Cape Floristic Region data}


Our data is extracted from a large database studying the distribution of plants in the Cape Floristic Region (CFR) of South Africa \citep{Takhtajan(86)}.  The CFR is one of the six floral kingdoms in the world and is located in the southwestern part of South Africa.  Though, geographically it is relatively small, it is extremely diverse ($9,000+$ species) and highly endemic ($70\%$ occur only in the CFR \citep{Rebelo(01)}).
There are more than $40,000$ sites with recorded sampling within the CFR.  The database from which our dataset was extracted consists of more than 1,400 plots with more than 2,800 species spanning six regions.
The data we use comes from one of these regions and exhibits high spatial clustering with $n=662$ plots and $S=639$ species. The response is binary, presence-absence for each species and plot (location).

\begin{figure}[ht]
  \begin{center}
   \includegraphics[width=13cm]{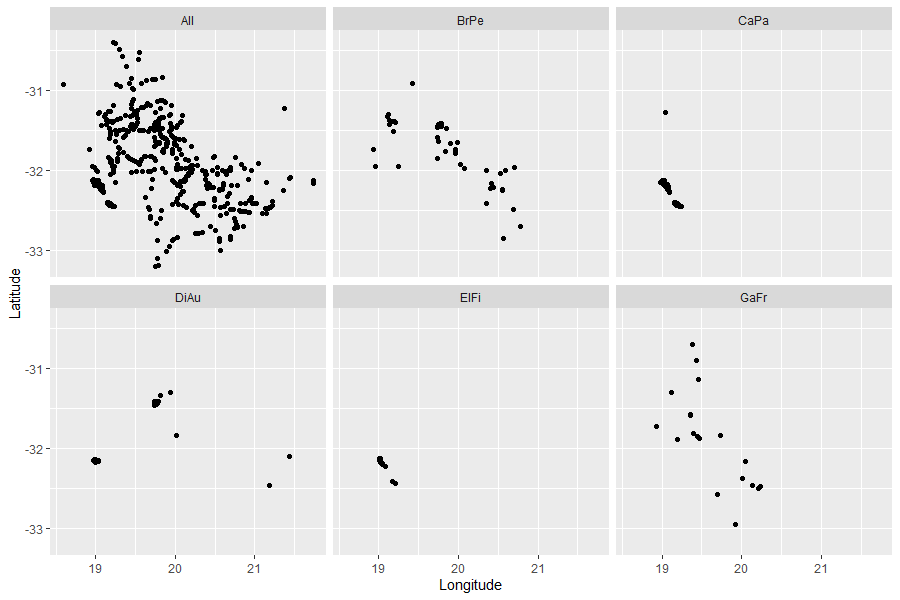}
  \end{center}
  \caption{All 662 locations in CFR and the distribution of the presence of selected 5 species.}
  \label{fig:CFR}
\end{figure}

The left panel of Figure \ref{fig:CFR} shows the 662 locations in CFR data and the right panel shows the distribution of 5 selected species: 1) {\it Bromus pectinatus} (BrPe); 2) {\it Cannomois parviflora} (CaPa); 3) {\it Diospyros austro-africana} (DiAu); 4) {\it Elegia filacea} (ElFi); and 5) {\it Galenia fruticosa} (GaFr). 

\begin{figure}[ht]
  \begin{center}
   \includegraphics[width=13cm]{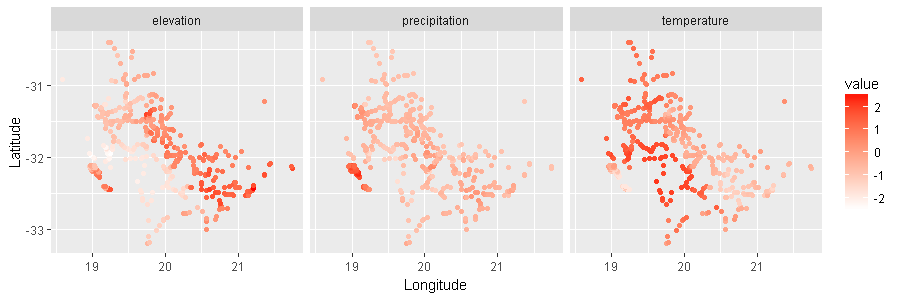}
  \end{center}
  \caption{Standardized covariate surface: elevation (left), mean annual precipitation (middle) and mean annual temperature (right)}
  \label{fig:CFRcov}
\end{figure}

As covariate information, we include: (1) elevation, (2) mean annual precipitation, and (3) mean annual temperature; these values are standardized, following \cite{Shirotaetal(19)}.
Figure \ref{fig:CFRcov} shows the three standardized covariate surfaces.

Altogether, the total number of binary responses is $n\times S=662\times 639=423,018$. The overall number of presences is 6,980, 1.65$\%$ of the total number of binary responses.  This emphasizes the fact that, although we have many species in our dataset, only a few are present on any given plot.

\begin{figure}[ht]
  \begin{center}
   \includegraphics[width=16cm]{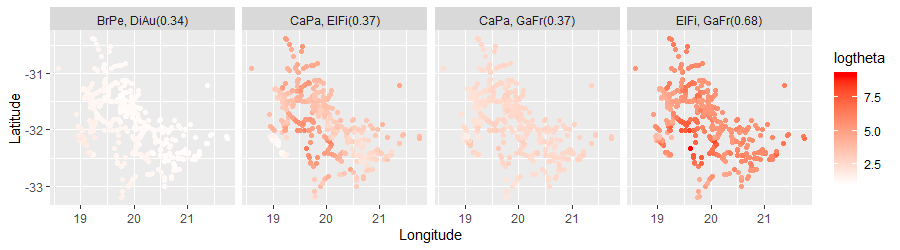}
  \end{center}
  \caption{Posterior mean surfaces of log odds ratio for all pairs of 5 species with positive correlation. The values in parenthesis are posterior correlation for each pair.}
  \label{fig:logtheta_10sp_pos}
\end{figure}

\begin{figure}[ht]
  \begin{center}
   \includegraphics[width=16cm]{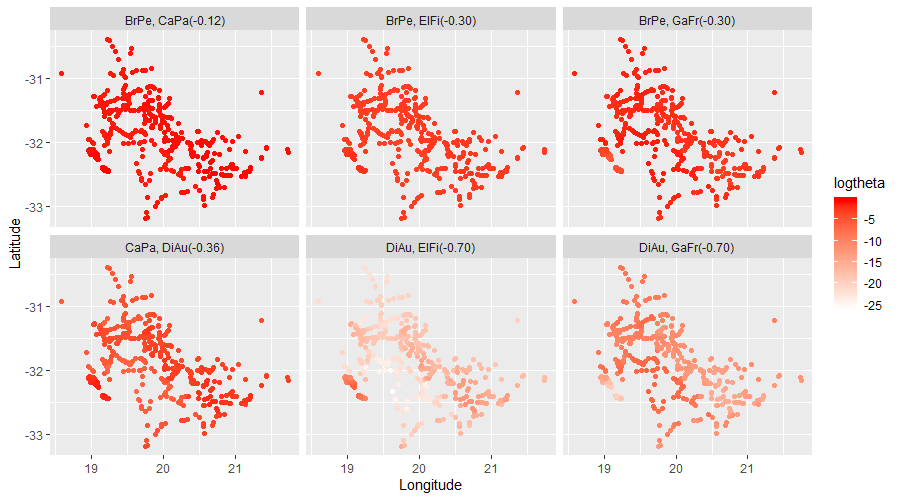}
  \end{center}
  \caption{Posterior mean surfaces of log odds ratio for all pairs of 5 species with negative correlation. The values in parenthesis are posterior correlation for each pair.}
  \label{fig:logtheta_10sp_neg}
\end{figure}

As noted above, it is difficult to grasp a $639 \times 639$ correlation matrix.  Furthermore, investigation of the resulting potential $639 \times 639/2 = 20,416$ odds ratio surfaces resulting from all possible pairs is infeasible.  So, we focus on the foregoing $5$ species which occur in at least $20$ of the locations and provide a range of correlations from $-.7$ to $.7$ under the model fitting.
The associated set of $10$ posterior log odds ratio ($\log\theta$) surfaces is presented in Figures \ref{fig:logtheta_10sp_pos} and \ref{fig:logtheta_10sp_neg}. Generally, $\log\theta$ is larger than 0 for positively correlated pairs and smaller than 0 for negatively correlated pairs as argued above. The probability of joint occurrence ($p_{11}$) is basically very small (close to 0) for all pairs and most locations, though there are some locations where $p_{11}$ reaches $0.15$ for positively correlated pairs.
The correlations of ElFi with the other species are quite different: -0.30 (BrPe), 0.37 (CaPa), -0.70 (DiAu) and 0.68 (GaFr).
For the species that are positively correlated with ElFi, i.e., CaPa and GaFr, large positive $\log \theta(\mathbf{s})$ is observed in the western part of the region. On the other hand, for the species that are negatively correlated with ElFi, i.e., BrPe and DiAu, the $\log \theta(\mathbf{s})$ surface in the western part of the region shows smaller negative values.
Additionally, the strongly negatively correlated pairs, (DiAu, ElFi) and (DiAu, GaFr) show negative $\log \theta(\mathbf{s})$ (minimum -26.1) in the middle eastern part of the region.



In summary, because we consider many species in our model but only a few occur at a site, the probability of joint absence is almost always very high. 
Since individual species modeling ignores all other species, $P(Y_{j}(\bs) =1)$ need not be very small.  The added insight from the current JSDMs is that while a marginal model may suggest that the chance of absence need not be high, the joint chance of absence will almost always be high.


\section{Odds ratios for data with ordinal abundance classification}

Suppose we have data for each species at each site in the form of an abundance.  When abundances are recorded as counts, they are often converted to ordinal classifications, e.g., (i) absent, (ii) $1$ to $5$, (iii) $6$ to $20$, (iv) $21$ to $50$, and (v) more than $50$.  In some cases, abundances are presented as \emph{relative} abundances, normalizing individual species abundance by total abundance across all species.  Now, we can specify ordinal classifications through proportions.  In other cases, abundances are viewed as cover-abundance, i.e., a measure of plant cover  (often used in vegetation science). It is based on percentages with several scales of cover-abundance in the literature, e.g. the 5-point cover scale of Braun-Blanquet or the Domin scale \citep{Mueller-DomboisEllenberg(74)}.

In any case, for a pair of species, we replace the foregoing $2 \times 2$ table with a $K \times K$ table if there are $K$ ordinal classifications, letting $Y_{i,j}$ denote the categorical classification for species $j$ at site $i$, taking ordered values $k=1,2,...,K$.  Then, we can consider local, global, and cumulative odds ratios \citep{Agresti(12)} as measures of species dependence.
They are defined as:
\begin{itemize}
    \item[(i)] \textbf{Local}: $\theta_{i,kk'}^{L(j,j')}= \frac{P(Y_{i,j}=k, Y_{i,j'} = k')P(Y_{i,j}=k+1, Y_{i,j'} = k'+1)}{P(Y_{i,j}=k, Y_{i,j'} = k'+1)P(Y_{i,j}=k+1, Y_{i,j'} = k')}$
    \item[(ii)] \textbf{Global}: $\theta_{i,kk'}^{G(j,j')}=\frac{P(Y_{i,j} \leq k, Y_{i,j'} \leq k')P(Y_{i,j} >k, Y_{i,j'} > k')}{P(Y_{i,j} \leq k, Y_{i,j'} > k')P(Y_{i,j} > k, Y_{i,j'} \leq k')}$
    \item[(iii)] \textbf{Cumulative}: $\theta_{i,kk'}^{C(j,j')}=\frac{P(Y_{i,j'} \leq k' | Y_{i,j} =k)/P(Y_{i,j'} >k' | Y_{i,j} = k)}{P(Y_{i,j'} \leq k' | Y_{i,j} = k+1)P(Y_{i,j'} > k' | Y_{i,j} = k+1)}$
\end{itemize}
By extending the discussion in Section 2, each of these odds ratios has a clear interpretation in terms of a particular type of pairwise species dependence.  The cumulative odds ratios are typically used as the basis for specifying ordinal categorical data models \citep{Agresti(12)}.  For example, using cumulative odds ratios, we can specify cumulative logit or probit models.  Such models are not our intention here.  Rather, in our setting, upon fitting a JSDM for ordinal abundances \citep[say using GJAM as in][]{Clarketal(17)}, we can use posterior samples from the latent Gaussian variables to obtain posterior distributions for any of the foregoing odds ratios for any pair of species at any site.  We note that GJAM fits the JSDM treating the sites as independent; at present, there is no spatial version for a JSDM with count data.  In any event, for the nonspatial case, we omit details and exemplification here, leaving this for future work.

\section{Conclusion}

This paper has attempted to bring some further light to the nature of joint species dependence that is captured by the current collection of joint species distribution models.  In particular, focusing on presence/absence data, we have argued that the correlations which are incorporated in the latent multivariate normal models that drive these JSDMs offer little useful inference and that odds ratios as well as joint occurrence probabilities, all induced by the JSDM specification, enable more clear interpretation.  Further, we have employed the extension of the JSDMs to spatial modeling of presence/absence in order to develop odds ratio surfaces as well as joint occurrence surfaces over a study region.  In this way, we can assess how joint species dependence varies over the region.

Because this contribution is primarily methodological we have not devoted substantial effort to the data analysis.  However, according to the application, more detailed dependence stories can be extracted.  In this regard, even if one fits a packaged JSDM  model, rather than a spatial JSDM, to a dataset with spatially varying regressors, we will still obtain spatially varying odds ratios and joint presence probabilities.  So, again, interpretation extends beyond the pairwise correlations.  Lastly, for abundance data as well as for compositional data, at present there are no spatial JSDMs.  A path for future research emerges.

\appendix
\section{Appendix}

Here, we consider more carefully the connection between the correlation arising under the latent multivariate normal model for species pairs and the associated odds ratio.  We draw on some older work relating bivariate normal probabilities to the associated bivariate correlation.  There is a substantial literature, with multivariate extensions, and we only note two papers here: \cite{Gupta(63)} and \cite{Slepian(62)}.

The basic result we need is the following:

Theorem: Suppose $\left(
               \begin{array}{c}
                 Z_{1} \\
                 Z_{2} \\
               \end{array}
                \right)$

 $\sim$ BivN$(\left(
             \begin{array}{c}
               0 \\
               0 \\
             \end{array}
           \right)$, $\left(
                      \begin{array}{cc}
                        1 & \rho \\
                        \rho & 1\\
                      \end{array}
                    \right))$.  Then, $P(Z_{1} \leq c_1, Z_{2} \leq c_2)$ is non-decreasing in $\rho$.

Now, let's apply this result to \eqref{eq:odds}.  For fixed $\mu_{i}^{(j)}$ and $\mu_{i}^{(j')}$, by simple probability calculations, we have $p_{i,00}^{(j,j')}$ non-decreasing in $\rho^{(j,j')}$,  we have $p_{i,01}^{(j,j')}$ non-increasing in $\rho^{(j,j')}$, we have $p_{i,10}^{(j,j')}$ non-increasing in $\rho^{(j,j')}$, and we have $p_{i,11}^{(j,j')}$ non-decreasing in $\rho^{(j,j')}$.  As a result, the numerator in \eqref{eq:odds} is non-decreasing in $\rho^{(j,j')}$ while the denominator in \eqref{eq:odds} is non-increasing in $\rho^{(j,j')}$.  So, altogether, we have $\theta_{i}^{(j,j')}$ non-decreasing in $\rho^{(j,j')}$ for all $i$ and $(j,j')$ pairs.

As a corollary, since $\theta_{i}^{(j,j')} =1$ when $\rho^{(j,j')} = 0$, we must have $\theta_{i}^{(j,j')} \geq 1$ when $\rho^{(j,j')} > 0$ and $\theta_{i}^{(j,j')} \leq 1$ when $\rho^{(j,j')} < 0$.

\bibliographystyle{chicago}
\bibliography{SP}

\end{document}